\documentclass[10pt, aps,prd,superscriptaddress,tightenlines,notitlepage,nofootinbib,floatfix]{revtex4-1}

\usepackage{multirow}
\usepackage{bigdelim}
\usepackage{graphicx}
\usepackage{color}
\usepackage{dcolumn}
\usepackage{physics}
\usepackage{mathtools}
\usepackage{bm}
\usepackage{array}
\usepackage{comment}
\usepackage{cancel}
\usepackage{here}
\usepackage[usenames,dvipsnames]{xcolor}
\usepackage{mathrsfs}
\usepackage{amsmath,amsthm}
\usepackage{enumerate}
\usepackage{empheq} 
\theoremstyle{definition}

\usepackage{xcolor} 
\definecolor{RoyalBlue}{rgb}{0.255, 0.412, 0.882}
\definecolor{DeepSkyBlue}{rgb}{0.0, 0.749, 1.0}
\definecolor{Crimson}{rgb}{0.862, 0.078, 0.235}
\definecolor{ForestGreen}{rgb}{0.133, 0.545, 0.133}
\definecolor{OrangeRed}{rgb}{1.0, 0.271, 0.0}
\definecolor{Orchid}{rgb}{0.855, 0.439, 0.839}
\definecolor{Sienna}{rgb}{0.627, 0.322, 0.176}
\definecolor{Goldenrod}{rgb}{0.855, 0.647, 0.125}
\definecolor{CadetBlue}{rgb}{0.372, 0.619, 0.627}
\definecolor{CornflowerBlue}{rgb}{0.392, 0.584, 0.929}
\definecolor{RebeccaPurple}{rgb}{0.4, 0.2, 0.6}
\definecolor{Salmon}{rgb}{0.980, 0.502, 0.447}
\definecolor{HotPink}{rgb}{1.0, 0.412, 0.706}
\definecolor{Chocolate}{rgb}{0.824, 0.412, 0.118}
\definecolor{SteelBlue}{rgb}{0.275, 0.510, 0.706}
\definecolor{FireBrick}{rgb}{0.698, 0.133, 0.133}
\definecolor{bondiblue}{rgb}{0.0, 0.58, 0.71}
\definecolor{celestialblue}{rgb}{0.29, 0.59, 0.82}
\definecolor{coolblack}{rgb}{0.0, 0.18, 0.39}
\definecolor{frenchblue}{rgb}{0.0, 0.45, 0.73}
\definecolor{lapislazuli}{rgb}{0.15, 0.38, 0.61}
\definecolor{mediumpersianblue}{rgb}{0.0, 0.4, 0.65}
\definecolor{darkpowderblue}{rgb}{0.0, 0.2, 0.6}
\definecolor{darkcandyapplered}{rgb}{0.64, 0.0, 0.0}
\definecolor{darkscarlet}{rgb}{0.34, 0.01, 0.1}
\definecolor{falured}{rgb}{0.5, 0.09, 0.09}
\definecolor{darkcyan}{rgb}{0.0, 0.55, 0.55}

\usepackage[pdftex,colorlinks,urlcolor=gray,citecolor=blue,linkcolor=blue]{hyperref}

\newcommand{\s}{\widetilde{S}}
\newcommand{\p}{\widetilde{P}}
\newcommand{\w}{\widetilde{W}}

\newcommand{\so}{S}
\newcommand{\po}{P}

\newcommand{\paren}[1]{\left(#1\right)}
\newcommand{\bck}[1]{\left[#1\right]}
\newcommand{\bce}[1]{\left\{#1\right\}}

\newcommand{\beq}{\begin{equation}}
\newcommand{\eeq}{\end{equation}}

\usepackage{amssymb}
\newcommand{\Rep}{\mathrm{Rep}}
\newcommand{\CA}{\mathrm{CA}_3}
\newcommand{\Vect}{\mathrm{Vec}}
\newcommand{\lprod}{\mathop{\overleftarrow{\prod}}}
\newcommand{\rc}{\check{R}}

\begin{document}

\author{Yuan Miao}
\affiliation{Laboratoire de Physique de l’Ecole Normale Supérieure, ENS, Université PSL, CNRS, Sorbonne Université, Université Paris Cité, Paris, France}

\author{Kohei Fukai}
\affiliation{RIKEN iTHEMS, Wako, Saitama 351-0198, Japan}

\date{\today}

\title{A Novel Algebraic Approach to Medium-Range Solvable Models}
\begin{abstract}
We extend Jones' Baxterisation, which produces solutions to the Yang-Baxter equation directly from the generators of an algebra, to quantum spin chains with medium-range interactions. 
The key ingredient is an extended Yang-Baxter equation for R matrices that are built from the generators attached to two neighbouring sites. Together with the accompanying Lax operator, it gives rise to transfer matrices that are in involution up to boundary terms, exactly as in Jones' construction. 
We find a solution to the extended Yang-Baxter equation, of non-difference form, in terms of the generators of the chromatic algebra of 3-colouring, and obtain from it two families of medium-range lattice Hamiltonians. Thus, each representation of the algebra provides a medium-range solvable model. 
Using the relation between the chromatic algebra of 3-colouring and the fusion category $\Rep(S_3)$, we construct four such representations. They reproduce the folded XXZ model and the hard-rod deformed XXZ model, and give new medium-range models of a three-state antiferromagnet and of a Rydberg-blockade ladder. Two of the four models possess non-invertible symmetries.

\end{abstract}

\maketitle

\section{Introduction}
\label{sec:intro}

Integrable quantum spin chains are a family of exactly solvable models that are defined on a one-dimensional array of quantum spins~\cite{Baxter1982book, Gaudin_2014, korepin_bogoliubov_izergin_1993, Takahashi1999book, sutherland2004beautiful}. Their history goes back to Bethe's solution of the spin-$\frac{1}{2}$ Heisenberg chain~\cite{Bethe1931}, followed by the analysis of the anisotropic XXZ chain by Yang and Yang~\cite{CNYang-anisotropic1-1966, CNYang-anisotropic2-1966} and the solution of the fully anisotropic XYZ chain by Baxter~\cite{Baxter1971prl, PhysRevLett.26.834, Baxter1973-i, Baxter1973-ii, Baxter1973-iii}. The solvability of these models is rooted in the Yang-Baxter equation~\cite{Yang1967, Baxter1972-eightvertex}. Its solution, the R matrix, is the building block of a family of mutually commuting transfer matrices, which is the starting point of the quantum inverse scattering method and the algebraic Bethe ansatz~\cite{Sklyanin1979, Takhtadzhan1979, korepin_bogoliubov_izergin_1993, Faddeev1996}. The logarithmic derivatives of the transfer matrix give rise to an extensive number of local conserved charges~\cite{Faddeev1996}.

Integrable models are not only of mathematical interest. Integrable spin chains describe quasi-one-dimensional magnetic materials, in which, for instance, the fractional spinon excitations of the Heisenberg chain have been observed by neutron scattering~\cite{Lake2005, Mourigal2013}. Integrable models have also been realised with ultracold atoms~\cite{RevModPhys.80.885, Kinoshita2006}. In the past two decades, integrable models have played an important role in the study of out-of-equilibrium properties of isolated quantum many-body systems~\cite{RevModPhys.83.863, CalabreseEsslerMussardo2016}. Due to the extensive number of conserved charges, an integrable system does not thermalise after a quantum quench, but relaxes to a generalised Gibbs ensemble~\cite{PhysRevLett.98.050405, Vidmar2016, doi:10.1126/science.1257026}, which in interacting spin chains has to include quasi-local~\cite{Wouters2014, PhysRevLett.113.117203, PhysRevLett.115.157201, Ilievski_2016} and non-commuting~\cite{PhysRevResearch.2.033403} conserved charges. The relaxation dynamics can be studied exactly, e.g.\ by means of the quench action~\cite{Caux2013quench, Caux:2016esd, EsslerFagotti2016}. On larger scales, the dynamics of integrable systems is described by generalised hydrodynamics~\cite{CastroAlvaredo2016, Bertini2016, Doyon2020}, whose predictions have been confirmed experimentally~\cite{Schemmer2019, Malvania2021}. Beyond condensed matter physics, the spectra of integrable spin chains are crucial in the study of the AdS/CFT correspondence~\cite{Minahan2003, Beisert2012}.

Since the R matrix is the basic building block of the transfer matrices, constructing integrable quantum spin chains amounts to finding solutions to the Yang-Baxter equation. Through the prism of quantum groups~\cite{Drinfeld1987, Jimbo1985}, R matrices can be obtained systematically by specifying the representation of the underlying quantum group~\cite{Bazhanov1985, Jimbo1986, jimbo-algebraic-book}. Meanwhile, Jones offered a different approach, called \emph{Baxterisation}~\cite{baxterization-1990}, obtaining solutions to the Yang-Baxter equation using algebras directly, such as the Temperley-Lieb algebra~\cite{temperley-lieb}, the Hecke algebra and the Birman-Wenzl-Murakami (BWM) algebra~\cite{BirmanWenzl1989, Murakami1987}. The approach has since been extended to other algebras such as the multi-colour braid-monoid algebras~\cite{GrimmPearce1993}. It has also been formulated in terms of braided tensor categories~\cite{Fendley2021braided} and applied to integrable quantum circuits~\cite{Miao2024floquet}. The virtue of this approach is that one solution in the algebra yields an integrable model for every representation of the algebra. For instance, the representations of the Temperley-Lieb algebra give the spin-$\frac{1}{2}$ XXZ chain, the $Q$-state Potts models and the restricted solid-on-solid (RSOS) models~\cite{Pasquier1988, pasquier1990common, martin1991potts, Nichols_2006}, as well as chains of interacting anyons~\cite{golden-chain}. Moreover, properties that are common to all the representations, such as the degeneracies of the spectra and the correlation functions, can be studied at the level of the algebra~\cite{klumper-TL-2010, fukai-TL-correlation-2023, fukai-current-minors-2026}.

In addition, medium-range spin chains have recently drawn attention, due to the close relation to Rydberg atom experiments~\cite{Bernien2017, Browaeys2020} and phenomena such as quantum many-body scars~\cite{Turner2018, Serbyn2021, Moudgalya2022review} and Hilbert space fragmentation~\cite{Sala2020, Khemani2020, Hilbert-Space-Fragmentation-2020, fragm-commutant-1}. Loosely speaking, medium-range spin chains stand for the spin chains with interaction beyond the nearest-neighbour one. Meanwhile, integrable medium-range spin chains are much less understood than their nearest-neighbour counterparts. Well-known integrable examples are the Bariev model~\cite{Bariev1991-original, Bariev1991-XY, Bariev1993}, the folded XXZ model~\cite{Lenart-folded1-2021, Lenart-folded2-2021, Pozsgay-folded-xxz-2021}, and the constrained models with Rydberg blockade~\cite{fendley-sachdev-rsos, AlcarazBariev1999, Corcoran2025rydberg} together with the related RSOS quantum chains~\cite{RSOS-H}, which descend from Baxter's hard-hexagon model and the RSOS models of Andrews, Baxter and Forrester~\cite{Baxter1980hardhexagon, ABF1984}. There are proposals to write down the Lax operators and the R matrices for the integrable medium-range spin chains~\cite{Pozsgay-medium-2021}, but only few explicit solutions are known~\cite{Pozsgay-medium-2021, Pozsgay-hardrod-2021, Corcoran2025rydberg}. For free-fermionic models, on the other hand, an algebraic approach to medium-range spin chains has proven very successful: the free fermions in disguise~\cite{disguise-Fendley-2019, Elman2021, disguise-generalized-2024, disguise-circuits-2024} are solvable in any representation of the algebra generated by the local terms of the Hamiltonian, which commute or anticommute depending on their distance. We overcome the problem for interacting models by following Jones' Baxterisation approach, obtaining solutions to the medium-range Yang-Baxter equations from an algebraic perspective. Some of the known solutions turn out to be representations of the algebraic solution obtained below.

The starting point of this article is the observation that the medium-range Yang-Baxter equations of Ref.~\cite{Pozsgay-medium-2021} admit a formulation which makes no reference to a tensor-product Hilbert space. Instead of an R matrix $\rc_m$ attached to one site $m$ of an algebra, as in Jones' approach, we consider an R matrix $\rc_{m,m+1}$ attached to two neighbouring sites, and replace the Yang-Baxter equation by the \emph{extended Yang-Baxter equation}~\eqref{eq:EYBE}, which relates $\rc_{m,m+1}$ to $\rc_{m+2,m+3}$. We show that the extended Yang-Baxter equation, and its companion for the Lax operator, lead to transfer matrices that commute up to boundary terms, and thus to medium-range Hamiltonians, in the same way as for Jones' Baxterisation.

We then present an explicit solution within the chromatic algebra of 3-colouring, i.e.\ the chromatic algebra~\cite{Fendley2008loops, FendleyKrushkal2009, FendleyKrushkal2010} with $Q=3$ colours supplemented with a Jones-Wenzl-type relation. This is the algebra that underlies the spin-$1/2$ XXZ chain and its non-invertible relatives discovered by Eck and Fendley~\cite{Fendley-chromatix-xxz}, who Baxterised it in the usual (nearest-neighbour) sense. The R matrix we find is of non-difference form, and gives rise to two families of medium-range Hamiltonians, $H$ and $\tilde{H}(v_0)$, defined in Eqs.~\eqref{eq:ECA-Hamiltonian} and \eqref{eq:Htilde-CA}.

Finally, the chromatic algebra of 3-colouring is intimately related to the fusion category $\Rep(S_3)$: its generators are built from the three fusion channels of two copies of the two-dimensional irreducible representation of the symmetric group $S_3$. Each of the four indecomposable module categories of $\Rep(S_3)$ therefore furnishes a representation of the algebra, which is the origin of the four models of Ref.~\cite{Fendley-chromatix-xxz}. Applying this construction to our solution, we find four medium-range models. Two of them are known, namely the folded XXZ model~\cite{Lenart-folded1-2021, Lenart-folded2-2021, Pozsgay-folded-xxz-2021} (together with its interacting generalisation) and the hard-rod deformed XXZ model~\cite{Pozsgay-medium-2021, Pozsgay-hardrod-2021}, which contains a special point of the Bariev model. The other two, defined on a three-state antiferromagnet and on a Rydberg-blockade ladder, are new to the best of our knowledge. The models are related to each other by non-invertible dualities, and two of them possess the non-invertible $\Rep(S_3)$ symmetry~\cite{BhardwajTachikawa2018}, and its continuous counterpart~\cite{YM_global_symm}.

The article is organised as follows. We review Jones' Baxterisation in Sec.~\ref{sec:jones}, with the Temperley-Lieb algebra and the chromatic algebra of 3-colouring as examples. The extended Yang-Baxter equation is introduced in Sec.~\ref{sec:proposal}, and its solution in the chromatic algebra of 3-colouring is given in Sec.~\ref{sec:solution}. The four representations are discussed in Sec.~\ref{sec:rep}. We conclude in Sec.~\ref{sec:summary}. The relations of the extended chromatic algebra are derived in Appendix~\ref{app:relations}.

\section{Baxterisation à la Jones}
\label{sec:jones}

For a given algebra $\mathcal{A}$ with generators $a^{(n)}_m$, labelled by the sites $m$ of a chain such that the generators at sites $m$ and $m'$ commute when $|m-m'| \geq 2$, one might find solution to the Yang-Baxter equation~(YBE)
\beq 
	\check{R}_m (u,v) \check{R}_{m+1} (w,v) \check{R}_m (w,u) = \check{R}_{m+1} (w,u) \check{R}_m (w,v) \check{R}_{m+1} (u,v) ,
	\label{eq:YBE1}
\eeq
where the R matrix is defined by the generators of $\mathcal{A}$,
\beq 
	\check{R}_m (u,v) = \sum_n f^{(n)} (u,v) a^{(n)}_m .
\eeq 

Once we find such a solution, it is straightforward to define transfer matrices
\beq 
	T(u,v_0 )  =  \lprod_{m=1}^{L} \check{R}_m (u,v_0) = \check{R}_L (u,v_0) \cdots \check{R}_2 (u,v_0) \check{R}_1 (u,v_0) ,
	\label{eq:T-jones}
\eeq 
which by using the YBE \eqref{eq:YBE1} recursively are in involution up to a boundary term,
\beq 
	\lim_{L\to \infty} \left[ T(u,v_0 ) , T(w,v_0 ) \right] = 0 .
	\label{eq:involution-jones}
\eeq 
To see this, we set $v = v_0$ in Eq.~\eqref{eq:YBE1} and rewrite it as
\beq 
	\check{R}_m (u,v_0) \check{R}_{m+1} (w,v_0) = \check{R}_{m+1} (w,u) \bck{ \check{R}_m (w,v_0) \check{R}_{m+1} (u,v_0) } \check{R}_{m}^{-1} (w,u) .
	\label{eq:exchange-jones}
\eeq 
Since $\check{R}_{m+1} (w,v_0)$ commutes with $\check{R}_{k} (u,v_0)$ for $k \leq m-1$, the product of two transfer matrices can be reordered as
\beq 
	T(u,v_0 ) T(w,v_0 ) = \check{R}_L (u,v_0) \bck{ \lprod_{m=1}^{L-1} \check{R}_m (u,v_0) \check{R}_{m+1} (w,v_0) } \check{R}_1 (w,v_0) .
	\label{eq:reorder-jones}
\eeq 
Applying Eq.~\eqref{eq:exchange-jones} to each factor, the operators $\check{R}_{m} (w,u)$ and $\check{R}_{m}^{-1} (w,u)$ cancel pairwise in the bulk, and we obtain
\beq 
	T(u,v_0 ) T(w,v_0 ) = \check{R}_L (u,v_0) \check{R}_L (w,u) \bck{ \lprod_{m=1}^{L-1} \check{R}_m (w,v_0) \check{R}_{m+1} (u,v_0) } \check{R}_{1}^{-1} (w,u) \check{R}_1 (w,v_0) ,
	\label{eq:involution-jones-finite}
\eeq 
which differs from $T(w,v_0 ) T(u,v_0 )$ only by operators acting at the two ends of the chain,
\beq
    T(w,v_0 ) T(u,v_0 ) = \check{R}_L (w,v_0) \check{R}_L^{-1} (w,u) \check{R}_L^{-1} (u,v_0) T(u,v_0 ) T(w,v_0 ) \check{R}_1^{-1} (w,v_0) \check{R}_1^{-1} (w,u) \check{R}_1 (u,v_0) .
\eeq
When the R matrix is regular, $\check{R}_m (v_0,v_0) = 1$, the logarithm of the transfer matrix generates local charges that commute with each other in the bulk. The first of them is the Hamiltonian
\beq 
	H = \sum_m \left. \partial_u \check{R}_m (u,v_0) \right|_{u=v_0} .
\eeq 

There are several renowned examples of the algebra $\mathcal{A}$ that possesses solution(s) to the YBE \eqref{eq:YBE1}, such as the Temperley-Lieb~(TL) algebra~\cite{temperley-lieb}, Birman-Wenzl-Murakami~(BWM) algebra~\cite{BirmanWenzl1989, Murakami1987} and more recently the chromatic algebra of 3-colouring~\cite{Fendley-chromatix-xxz}. We exemplify the usage of the Baxterisation procedure using the first and the last ones.

\textit{Temperley-Lieb algebra.---} 
The TL algebra is defined by the generators $e_m$ satisfying
\beq 
	e_m^2 = \beta e_m , \quad e_m e_{m \pm 1} e_m = e_m ,
\eeq 
\beq 
	\left[ e_m , e_n \right] = 0 , \quad |m-n | \geq 2 .
\eeq 
The fugacity parameter $\beta = q+q^{-1} = 2 \cosh \eta$.

The baxterised R matrix reads
\beq 
	\check{R}_m (u,v) = \check{R}_m (u-v) = 1 - \frac{\sinh \lambda }{\sinh(\lambda +\eta)} e_m , \quad \lambda = u-v ,
\eeq 
which is a solution to the YBE~\eqref{eq:YBE1}. The integrable Hamiltonian thus reads, up to a multiplicative constant,
\beq 
	H = \sum_m e_m .
\eeq 

There are numerous representations of the TL algebra, such as Ising model, Potts models, spin-$1/2$ XXZ model, and SOS models~\cite{Baxter1982book, Pasquier1988, Nichols_2006}.

\emph{Chromatic algebra of 3-colouring.---} 
We review the chromatic algebra of 3-colouring~($\CA$)~\cite{Fendley-chromatix-xxz}. 
The chromatic algebra~\cite{Fendley2008loops, FendleyKrushkal2009, FendleyKrushkal2010} is a diagrammatic algebra depending on a parameter $Q$, whose evaluation gives the chromatic polynomial, i.e. the number of $Q$-colourings of the regions of a planar graph. By $\CA$ we denote the chromatic algebra at $Q=3$ supplemented with a Jones--Wenzl-type relation, in the presentation of Ref.~\cite{Fendley-chromatix-xxz}.

$\CA$ is defined by the generators $\so_j$ and $\po_j$ with $j = 1, \ldots, L$. 
The relations of the generators with the same indices are
\begin{align}
    \label{eq:CA-def1}
    \so_j^2 = 1 - \po_j
    ,\quad 
    \po_j^2 = \po_j
    ,\quad 
    \so_j \po_j = \po_j \so_j = 0.
\end{align}
The relation involving nearest indices is
\begin{align}
    \label{eq:CA-def2}
    \so_j \so_{j \pm 1} \so_j 
    =
    \po_j \so_{j \pm 1} \po_j 
    =0, 
\end{align}
where the periodic boundary condition is assumed: $\so_{L+1} = \so_{1}$. The generators satisfy the commutation relations:
\begin{align}
    \label{eq:CA-def3}
    \begin{aligned}
        &\bck{ \so_j, \so_l }
        =
        \bck{ \so_j, \po_l }
        = 0
        ,\quad 
        ( |j-l| > 1 )
        \\
        &\bck{ \po_j, \po_l } 
        = 0
        .\quad 
        ( \forall j, l)
    \end{aligned}
\end{align}
A useful consequence of the relations~\eqref{eq:CA-def1} and \eqref{eq:CA-def2} is~\cite{Fendley-chromatix-xxz}
\begin{align}
    \label{eq:CA-PS}
    \po_j \so_{j \pm 1} = \so_{j \pm 1} \paren{ 1 - \po_j } .
\end{align}

The Hamiltonian of the chromatic algebra of 3-colouring is abstractly defined by:
\begin{align}
    \label{eq:CA-Hamiltonian}
    H^{\mathrm{CA}}
    =
    \sum_{j=1}^{L} \paren{ \so_j - \Delta \so_j^2 }
    ,
\end{align}
where $\Delta$ is the anisotropic parameter. The Hamiltonian~\eqref{eq:CA-Hamiltonian} is proven to be integrable by Baxterisation~\cite{Fendley-chromatix-xxz}. The baxterised R matrix is of difference form,
\begin{align}
    \label{eq:CA-R}
    \check{R}_j (u,v) = \check{R}_j (u-v)
    ,\quad 
    \check{R}_j (\lambda)
    =
    \frac{ \sin(\lambda + \mu) }{ \sin \mu }
    + 
    \frac{ \sin \lambda }{ \sin \mu } \so_{j} 
    + 
    \paren{ 1 - \frac{ \sin(\lambda+\mu) }{ \sin \mu } } \so_{j}^2
    ,
\end{align}
and solves the YBE~\eqref{eq:YBE1} for arbitrary $\mu$.

The Hamiltonian~\eqref{eq:CA-Hamiltonian} is recovered from $\sin \mu \, \partial_\lambda \check{R}_j (\lambda) |_{\lambda = 0} = \so_j - \Delta \so_j^2 + \cos \mu$ with $\Delta = \cos \mu$.

As an example, the spin-$\frac{1}{2}$ XXZ chain is one of the representations of $\CA$. 
The following representation satisfies the defining relation of $\CA$~\eqref{eq:CA-def1}--\eqref{eq:CA-def3}:
\begin{align}
    \label{eq:CA-rep-xxz}
    \so_j^{\mathrm{XXZ}} &=  \frac{1}{2} ( X_{j}X_{j+1} + Y_{j}Y_{j+1} )
    ,
    \\
    \po_j^{\mathrm{XXZ}} &= \frac{1}{2} ( 1 + Z_{j}Z_{j+1} )
    ,
    \label{eq:CA-rep-xxz-P}
\end{align}
where $X_j$, $Y_j$ and $Z_j$ are the Pauli matrices acting on the $j$-th spin. In this representation, $H^{\mathrm{CA}}$ becomes the Hamiltonian of the spin-$1/2$ XXZ chain with anisotropy $\Delta$, up to a constant. The other known representations are discussed in Sec.~\ref{sec:rep}.

\section{A New Proposal for Baxterisation}
\label{sec:proposal}

In order to cope with the integrability of the medium-range spin chains, we need to extend the algebraic construction of the Baxterisation procedure à la Jones~\cite{baxterization-1990}. Inspired by the medium-range Lax formalism of Ref.~\cite{Pozsgay-medium-2021}, we propose the following construction.

Let $\mathcal{A}$ be an algebra with elements $O^{(a)}_m$ labelled by the sites $m$ of a chain, such that
\beq 
	\bck{ O^{(a)}_m , O^{(b)}_{m'} } = 0 , \quad |m - m'| \geq 3 .
	\label{eq:locality}
\eeq 
Instead of an R matrix attached to a single site, we consider an R matrix attached to two neighbouring sites,
\beq 
	\check{R}_{m,m+1} (u,v) = \sum_{a,b} \bck{ c_{ab} (u,v) \, O^{(a)}_m O^{(b)}_{m+1} + c'_{ab} (u,v) \, O^{(b)}_{m+1} O^{(a)}_{m} } ,
	\label{eq:R-ansatz}
\eeq 
and we propose the following \emph{extended Yang-Baxter equation},
\beq 
\begin{split}
	& \check{R}_{m,m+1} (u,v) \check{R}_{m+2,m+3} (w,v) \check{R}_{m,m+1} (w,u) = \\
	& \check{R}_{m+2,m+3} (w,u) \check{R}_{m,m+1} (w,v) \check{R}_{m+2,m+3} (u,v) .
\end{split}
	\label{eq:EYBE}
\eeq 
In contrast to the YBE~\eqref{eq:YBE1}, the two R matrices entering Eq.~\eqref{eq:EYBE} do not share any site. The equation is nevertheless non-trivial, because the elements at the sites $m$ and $m+1$ do not commute with the ones at the sites $m+2$ and $m+3$, except for the pair $(m, m+3)$.

The extended YBE is accompanied by a \emph{Lax operator} attached to a single site,
\beq 
	\check{L}_{m} (u) = \sum_{a} f_{a} (u) \, O^{(a)}_m ,
	\label{eq:L-ansatz}
\eeq 
which is intertwined by the R matrix through the RLL relation
\beq 
	\check{R}_{m,m+1} (u,v) \check{L}_{m-1} (u) \check{L}_{m+1} (v) = \check{L}_{m-1} (v) \check{L}_{m+1} (u) \check{R}_{m-1,m} (u,v) .
	\label{eq:RLL}
\eeq 
We assume that both operators are regular,
\beq 
	\check{L}_{m} (0) = 1 , \quad \check{R}_{m,m+1} (u,u) = 1 .
	\label{eq:regularity}
\eeq 

Eqs.~\eqref{eq:EYBE} and \eqref{eq:RLL} are the algebraic counterparts of the relations satisfied by the Lax operator and the R matrix of the spin chains with three-site interactions in Ref.~\cite{Pozsgay-medium-2021}, to which they are equivalent in a representation where $O^{(a)}_m$ act on the sites $m-1$, $m$ and $m+1$ of a tensor-product Hilbert space. Here, however, no reference to a particular representation is made. In the representations of Sec.~\ref{sec:rep}, the elements $O^{(a)}_m$ act on three, four or five neighbouring sites, and the Hilbert space is not necessarily a tensor product.

\textit{Transfer matrices.---} 
From the Lax operator, we define the transfer matrix
\beq 
	T(u) = \lprod_{m} \check{L}_{m} (u) = \cdots \check{L}_{m+1} (u) \check{L}_{m} (u) \check{L}_{m-1} (u) \cdots .
	\label{eq:T-lax}
\eeq 
Due to Eq.~\eqref{eq:locality}, $\check{L}_{m+1} (w)$ commutes with $\check{L}_{k} (u)$ for $k \leq m-2$, and the product of two transfer matrices can be reordered as
\beq 
	T(u) T(w) = \lprod_{m} \bck{ \check{L}_{m-1} (u) \check{L}_{m+1} (w) } .
	\label{eq:reorder-lax}
\eeq 
The RLL relation~\eqref{eq:RLL} implies
\beq 
	\check{L}_{m-1} (u) \check{L}_{m+1} (w) = \check{R}_{m,m+1}^{-1} (u,w) \bck{ \check{L}_{m-1} (w) \check{L}_{m+1} (u) } \check{R}_{m-1,m} (u,w) ,
	\label{eq:exchange-lax}
\eeq 
so that the R matrices cancel pairwise between the neighbouring factors. We thus prove that the transfer matrices are in involution up to boundary terms,
\beq 
	T(u) T(w) \simeq T(w) T(u) ,
	\label{eq:involution-lax}
\eeq 
where $\simeq$ denotes the equality up to operators acting at the two ends of the chain, cf. Eq.~\eqref{eq:involution-jones}.

To make the boundary terms in Eq.~\eqref{eq:involution-lax} explicit, we consider a chain with the sites $m = 1, \ldots , L$, for which the transfer matrix~\eqref{eq:T-lax} is $T(u) = \check{L}_{L} (u) \cdots \check{L}_{2} (u) \check{L}_{1} (u)$. 
In the product $T(u) T(w)$, we move each $\check{L}_{m+1} (w)$ with $m \geq 2$ to the left, through the operators $\check{L}_{k} (u)$ with $k \leq m-2$ that it commutes with, until it stands next to $\check{L}_{m-1} (u)$. The reordering~\eqref{eq:reorder-lax} then reads
\beq 
	T(u) T(w) = \check{L}_{L} (u) \check{L}_{L-1} (u) \bck{ \lprod_{m=2}^{L-1} \check{L}_{m-1} (u) \check{L}_{m+1} (w) } \check{L}_{2} (w) \check{L}_{1} (w) ,
	\label{eq:reorder-lax-finite}
\eeq 
where only the two leftmost Lax operators with the spectral parameter $u$ and the two rightmost ones with $w$ are left unpaired.

Applying Eq.~\eqref{eq:exchange-lax} to each factor in the bracket, the factor with the label $m$ contributes $\check{R}_{m-1,m} (u,w)$ on its right, and the next factor, with the label $m-1$, contributes $\check{R}_{m-1,m}^{-1} (u,w)$ on its left. The R matrices thus cancel pairwise in the bulk, and only the leftmost and the rightmost ones survive,
\beq 
	T(u) T(w) = \check{L}_{L} (u) \check{L}_{L-1} (u) \check{R}_{L-1,L}^{-1} (u,w) \bck{ \lprod_{m=2}^{L-1} \check{L}_{m-1} (w) \check{L}_{m+1} (u) } \check{R}_{1,2} (u,w) \check{L}_{2} (w) \check{L}_{1} (w) .
	\label{eq:involution-lax-finite}
\eeq 
On the other hand, the reordering~\eqref{eq:reorder-lax-finite} with $u$ and $w$ exchanged gives
\beq 
	T(w) T(u) = \check{L}_{L} (w) \check{L}_{L-1} (w) \bck{ \lprod_{m=2}^{L-1} \check{L}_{m-1} (w) \check{L}_{m+1} (u) } \check{L}_{2} (u) \check{L}_{1} (u) .
	\label{eq:reorder-lax-finite-2}
\eeq 
The products~\eqref{eq:involution-lax-finite} and \eqref{eq:reorder-lax-finite-2} share the same bulk part, i.e. the bracket, and differ only by operators acting on the sites $L-1$, $L$ and $1$, $2$ at the two ends of the chain. This is the precise meaning of Eq.~\eqref{eq:involution-lax}, in complete analogy with Eqs.~\eqref{eq:reorder-jones} and \eqref{eq:involution-jones-finite} for Jones' Baxterisation.

A second family of transfer matrices is obtained from the R matrix itself. Let us denote $\check{\mathsf{R}}_n (u,v) = \check{R}_{2n,2n+1} (u,v)$. Due to Eq.~\eqref{eq:locality}, $\check{\mathsf{R}}_n$ and $\check{\mathsf{R}}_{n'}$ commute when $|n - n'| \geq 2$, while the extended YBE~\eqref{eq:EYBE} becomes
\beq 
	\check{\mathsf{R}}_n (u,v) \check{\mathsf{R}}_{n+1} (w,v) \check{\mathsf{R}}_n (w,u) = \check{\mathsf{R}}_{n+1} (w,u) \check{\mathsf{R}}_n (w,v) \check{\mathsf{R}}_{n+1} (u,v) ,
\eeq 
which is nothing but the YBE~\eqref{eq:YBE1}. The transfer matrix thus becomes
\beq 
	\tilde{T} (u , v_0) = \lprod_{n} \check{R}_{2n,2n+1} (u,v_0) ,
	\label{eq:T-R}
\eeq 
and the argument of Sec.~\ref{sec:jones} shows that the transfer matrices $\tilde{T} (u , v_0)$ with the same $v_0$ are in involution up to boundary terms, exactly like Jones' case.

\textit{Hamiltonians.---}
Thanks to the regularity~\eqref{eq:regularity}, the logarithms of the transfer matrices~\eqref{eq:T-lax} and \eqref{eq:T-R} are sums of local densities. At the leading order, we obtain the Hamiltonians
\begin{align}
	H &= \sum_m \left. \partial_u \check{L}_{m} (u) \right|_{u=0} ,
	\label{eq:H-general}
	\\
	\tilde{H} (v_0) &= \sum_n \left. \partial_u \check{R}_{2n,2n+1} (u,v_0) \right|_{u=v_0} .
	\label{eq:Htilde-general}
\end{align}
The Hamiltonian $H$ is invariant under the translation by one site, while $\tilde{H} (v_0)$ is invariant under the translation by two sites. For an R matrix of non-difference form, $\tilde{H} (v_0)$ depends non-trivially on the parameter $v_0$. The higher charges follow from the expansion of the logarithm of the transfer matrices. For instance, the next charge generated by $T(u)$ reads
\beq 
	Q^{(2)} = \frac{1}{2} \sum_m \paren{ \bck{ \check{L}_{m+1}' , \check{L}_{m}' } + \bck{ \check{L}_{m+2}' , \check{L}_{m}' } + \check{L}_{m}'' - ( \check{L}_{m}' )^2 } ,
	\label{eq:Q2-general}
\eeq 
where $\check{L}_{m}' = \partial_u \check{L}_{m} (u) |_{u=0}$ and $\check{L}_{m}'' = \partial_u^2 \check{L}_{m} (u) |_{u=0}$.

Since the transfer matrices commute up to boundary terms, the local charges commute with each other in the bulk. As in the case of Jones' Baxterisation, this leads us to conjecture that the Hamiltonians~\eqref{eq:H-general} and \eqref{eq:Htilde-general} are integrable when the periodic boundary condition is imposed, i.e. that they possess an extensive number of mutually commuting local charges, in any representation of $\mathcal{A}$. For the solution presented below, we have verified explicitly that the Hamiltonians commute with the first higher charges on periodic chains in all the representations of Sec.~\ref{sec:rep}.

\section{Example: Chromatic algebra of 3-colouring}
\label{sec:solution}

We now present a solution to the extended YBE~\eqref{eq:EYBE} and the RLL relation~\eqref{eq:RLL} in terms of the generators of $\CA$.

\textit{Extended chromatic algebra.---} 
We introduce the extended chromatic algebra~(ECA). ECA is the subalgebra of $\CA$ generated by the following elements $\s_{j}$, $\p_j$, and $\w_j$ with $j= 1, \ldots ,L$:
\begin{align}
    \label{eq:ECA-embedding}
    \begin{aligned}
        \s_{j} 
        &= 
        \frac{1}{2} \bck{ \bck{ \po_{j-1}, \so_{j}}, \po_{j+1} } + \frac{1}{2} \so_{j}
        = \paren{ \po_{j-1} - \po_{j+1} }^2 \so_j ,
        \\
        \w_j
        &=
        \frac{1}{2} \left[ 1 + (2\po_{j-1} - 1) (2\po_{j} - 1) (2\po_{j+1} - 1) \right],
        \\
        \p_j 
        &=
        \po_j \w_j
        .
    \end{aligned}    
\end{align}
The operator $( \po_{j-1} - \po_{j+1} )^2$ is a projector that commutes with $\so_j$. The generator $\s_j$ is thus the generator $\so_j$ \emph{controlled} by its neighbours: it acts as $\so_j$ when exactly one of $\po_{j-1}$ and $\po_{j+1}$ takes the value $1$, and vanishes otherwise. 
Using the relations of $\CA$~\eqref{eq:CA-def1}--\eqref{eq:CA-PS}, one shows that the generators of ECA satisfy the following relations, cf. Appendix~\ref{app:relations}. The relations of the generators with the same indices are
\begin{align}
    \label{eq:ECA-def-1}
    \begin{aligned}
        &
        \s_{j}^2 = \w_j - \p_j
        ,\quad
        \p_j^2 = \p_j
        ,\quad
        \w_j^2 = \w_j
        ,
        \\
        &
        \s_{j} \p_j = \p_j \s_{j} = 0,
        \\
        &
        \s_{j} \w_j = \w_j \s_{j} = \s_{j}
        .
    \end{aligned}
\end{align}
There are relations involving indices up to the next nearest neighbour:
\begin{align}
    \label{eq:ECA-def-2}
    A_j \s_{j \pm d} A_j 
    = 0
    .\quad
    ( A_j \in \{\s_{j}, \p_j, \w_j\},\ d \in  \{1,2\} )
\end{align}
Here we assume the periodic boundary condition: $\s_{L+d} = \s_{d}$. 
The generators satisfy the commutation relation:
\begin{align}
    \label{eq:ECA-def-3}
    \begin{aligned}
        &
        \bck{ A_j, B_l } 
        = 0
        .\quad
        ( A_j, B_j \in \{\p_j, \w_j\}, \forall j,l )
        \\
        &
        \bck{ \s_{j}, A_l } 
        = 0
        .\quad 
        ( A_j \in \{\s_{j}, \p_j, \w_j\},\ |j-l|>2 )
    \end{aligned}
\end{align}
In addition, the counterpart of Eq.~\eqref{eq:CA-PS} holds,
\begin{align}
    \label{eq:ECA-def-4}
    \w_j \s_{j \pm d} = \s_{j \pm d} \paren{ 1 - \w_j }
    ,\quad
    ( d \in  \{1,2\} )
\end{align}
which implies Eq.~\eqref{eq:ECA-def-2} with $A_j = \w_j$. Compared with Eq.~\eqref{eq:locality}, the generators of ECA with the index $j$ play the role of the elements $O^{(a)}_j$ of Sec.~\ref{sec:proposal}. 
Note that $\s_j^3 = \s_j$, i.e. $\frac{1}{2} ( \s_j^2 \pm \s_j )$ and $1 - \s_j^2$ are mutually orthogonal projectors. It is tempting to regard Eqs.~\eqref{eq:ECA-def-1}--\eqref{eq:ECA-def-4} as the defining relations of an abstract algebra, of which Eq.~\eqref{eq:ECA-embedding} provides a realisation in $\CA$. We do not pursue this direction here, and all the statements below are understood as identities in $\CA$.

A remarkable property of ECA is that all of its elements commute with
\begin{align}
    \label{eq:NP}
    N_{P} = \sum_{j=1}^{L} \po_j ,
\end{align}
which is not the case for $\CA$ itself. Indeed, $\so_j$ changes the values of both $\po_{j-1}$ and $\po_{j+1}$, cf. Eq.~\eqref{eq:CA-PS}, and the projector in $\s_j$ selects the processes in which $\po_{j-1} + \po_{j+1}$ is unchanged. Any Hamiltonian built from the generators of ECA therefore possesses a $U(1)$ symmetry generated by $N_P$, in addition to the symmetries of $\CA$.

\textit{Lax operator.---} 
We define the abstract Hamiltonian of the ECA by:
\begin{align}
    \label{eq:ECA-Hamiltonian}
    H^{\mathrm{ECA}} = \sum_{j=1}^{L} h_j 
    ,\quad
    h_j = \s_{j} - \Delta \s_j^2
    = \paren{ \po_{j-1} - \po_{j+1} }^2 \paren{ \so_j + \Delta \so_j^2 } 
    .
\end{align}
We prove that the Hamiltonian of ECA~\eqref{eq:ECA-Hamiltonian} is of the form~\eqref{eq:H-general}, i.e. that it follows from a solution to the RLL relation~\eqref{eq:RLL}.

We define the following Lax operator,
\begin{align}
    \check{L}_{j}(u;\mu)
    = 
    \label{eq:ECA-lax}
    \frac{ \sin(u + \mu) }{ \sin \mu }
    + 
    \frac{ \sin u }{ \sin \mu } \s_{j} 
    + 
    \paren{ 1 - \frac{ \sin(u+\mu) }{ \sin \mu } } \s_{j}^2
    .
\end{align}
It is obtained from the R matrix of $\CA$~\eqref{eq:CA-R} by replacing $\so_j$ with $\s_j$. Equivalently, the Lax operator acts as the R matrix~\eqref{eq:CA-R} when exactly one of $\po_{j-1}$ and $\po_{j+1}$ takes the value $1$, and as a scalar otherwise. 
It satisfies $\check{L}_{j}(0;\mu) = 1$ and the inversion relation
\begin{align}
    \check{L}_{j}(u;\mu) \check{L}_{j}(-u;\mu) = \frac{ \sin (\mu + u) \sin (\mu - u) }{ \sin^2 \mu } .
    \label{eq:ECA-inversion}
\end{align}
We construct the transfer matrix from the Lax operator:
\begin{align}
    \label{eq:ECA-transfer-mat}
    T(u;\mu)
    =
    \lprod_{j} \check{L}_{j}(u;\mu)
    ,
\end{align}
where we assume the $L \rightarrow \infty$.

The Hamiltonian of ECA is recovered from the first derivative of the transfer matrix, which can be seen from
\begin{align}
    \left. \partial_{u} \check{L}_{j}(u;\mu) \right|_{u = 0} 
    =
    \frac{1}{\sin \mu} \paren{ \s_j - \Delta \s_j^2 + \cos \mu }
    ,
\end{align}
where $\Delta = \cos \mu$. The regime $|\Delta| > 1$ is reached by considering imaginary values of $\mu$.

\textit{R matrix.---}
We also define the R-matrix by:
\begin{widetext}
\begin{align}
\label{eq:ECA-R-matrix}
\begin{aligned}
\check{R}_{j,j+1}(u,v;\mu)
=&
1 
+ 
\paren{ \frac{ \sin \mu }{ \sin(\lambda+\mu) } - 1 } \paren{ \s_{j}^2 + \s_{j+1}^2 } 
+
\frac{ \sin \lambda }{ \sin(\lambda+\mu) } \paren{ \bce{ \p_{j}, \s_{j+1}} + \bce{ \p_{j+1}, \s_{j} } }
\\
&
+ 
\bck{ 1 - \frac{ 2 \sin \mu }{ \sin(\lambda+\mu) } + \frac{1}{ \sin(v-\mu) \sin(u+\mu) } \paren{\sin u \sin v - \frac{ \sin^3 \mu }{ \sin(\lambda+\mu) } } }  \s_{j}^2 \s_{j+1}^2 
\\
& 
+ 
\frac{ \sin \lambda }{ \sin(\lambda+\mu) } \paren{ \frac{ \sin u }{ \sin(u+\mu) } \s_{j+1} \s_{j} + \frac{ \sin v }{ \sin(v-\mu) } \s_{j} \s_{j+1} }
\\
& 
+
\frac{ \sin \lambda \sin \mu }{ \sin(\lambda+\mu) } \bck{ \frac{ \s_{j+1} \s_{j}^2 + \s_{j+1}^2 \s_{j} }{ \sin(u+\mu) } - \frac{ \s_{j} \s_{j+1}^2 + \s_{j}^2 \s_{j+1} }{ \sin(v-\mu) } }
,
\end{aligned}
\end{align}
\end{widetext}
where $\lambda = u - v$ and $\{ A, B \} = AB + BA$ denotes the anticommutator. 
The R matrix is of the form~\eqref{eq:R-ansatz}. In contrast to the R matrix of $\CA$~\eqref{eq:CA-R}, it is \emph{not} of difference form, similar to the R matrix of the Bariev model~\cite{Zhou1996, Shiroishi1997}.

Then, we show the following RLL relation holds~\cite{Pozsgay-medium-2021}, following from the defining properties~\eqref{eq:ECA-def-1}, \eqref{eq:ECA-def-2} and \eqref{eq:ECA-def-3},
\begin{align}
    \label{eq:ECA-RLL}
    \begin{aligned}
        &
        \check{R}_{j,j+1}(u,v;\mu) \check{L}_{j-1}(u;\mu) \check{L}_{j+1}(v;\mu)
        \\
        & \hspace{2em}
        =
        \check{L}_{j-1}(v;\mu) \check{L}_{j+1}(u;\mu) \check{R}_{j-1,j}(u,v;\mu)
        ,
    \end{aligned}
\end{align}
which ensures the mutual commutativity of the transfer matrix~\eqref{eq:ECA-transfer-mat} up to boundary terms, $T(u;\mu) T(v;\mu) \simeq T(v;\mu) T(u;\mu)$, cf. Eq.~\eqref{eq:involution-lax}.

Furthermore, the R-matrix itself satisfies the extended Yang-Baxter equation~\eqref{eq:EYBE}:
\begin{align}
    \label{eq:ECA-YBE}
    \begin{aligned}
        &
        \check{R}_{j-1,j}(u,v;\mu) \check{R}_{j+1,j+2}(w,v;\mu) \check{R}_{j-1,j}(w,u;\mu)
        \\
        & \hspace{2em}
        =
        \check{R}_{j+1,j+2}(w,u;\mu) \check{R}_{j-1,j}(w,v;\mu) \check{R}_{j+1,j+2}(u,v;\mu)
        .
    \end{aligned}
\end{align}
The R matrix is regular and unitary,
\begin{align}
    \label{eq:ECA-unitarity}
    \check{R}_{j,j+1}(u,u;\mu) = 1
    ,\quad
    \check{R}_{j,j+1}(u,v;\mu) \check{R}_{j,j+1}(v,u;\mu) = 1
    ,
\end{align}
and it factorises into a product of two Lax operators when the second spectral parameter vanishes,
\begin{align}
    \label{eq:ECA-factorisation}
    \check{R}_{j,j+1}(u,0;\mu) = \frac{ \sin^2 \mu }{ \sin^2 (u + \mu) } \, \check{L}_{j+1}(u;\mu) \check{L}_{j}(u;\mu)
    ,
\end{align}
in agreement with the general structure found in Ref.~\cite{Pozsgay-medium-2021}.

Eqs.~\eqref{eq:ECA-RLL}--\eqref{eq:ECA-factorisation} are identities in $\CA$, i.e. they follow from the relations~\eqref{eq:CA-def1}--\eqref{eq:CA-def3} alone, and thus hold in any representation of $\CA$. Since the direct verification is tedious, we have checked them with the help of computer algebra with arbitrary spectral parameters and $\mu$.

The higher charges follow from Eq.~\eqref{eq:Q2-general}. Due to the inversion relation~\eqref{eq:ECA-inversion}, the last two terms in Eq.~\eqref{eq:Q2-general} add up to a constant, and the first higher charge takes a simple form,
\begin{align}
    \label{eq:ECA-Q2}
    Q^{(2)} \propto \sum_{j=1}^{L} \bck{ h_j , h_{j+1} + h_{j+2} } .
\end{align}
We have verified that $\bck{ H^{\mathrm{ECA}} , Q^{(2)} } = 0$ on periodic chains in all the representations of Sec.~\ref{sec:rep}.

\textit{Second family of Hamiltonians.---}
We now turn to the Hamiltonian~\eqref{eq:Htilde-general} obtained from the R matrix~\eqref{eq:ECA-R-matrix}. With a convenient normalisation,
\begin{align}
    \label{eq:Htilde-CA}
    \tilde{H}^{\mathrm{ECA}} (v_0) 
    = 
    \sin \mu \sum_{n=1}^{L/2} \left. \partial_u \check{R}_{2n,2n+1} (u,v_0;\mu) \right|_{u=v_0}
    =
    H^{\mathrm{ECA}} + \sum_{n=1}^{L/2} \delta h_{2n,2n+1} (v_0)
    ,
\end{align}
where $L$ is even, and we used Eq.~\eqref{eq:ECA-def-4} to simplify the result. The deformation acts on every other pair of neighbouring sites,
\begin{align}
    \label{eq:delta-h}
    \begin{aligned}
    \delta h_{j,j+1} (v_0) 
    =& \,
    \alpha_+ \s_{j+1} \s_{j} + \alpha_- \s_{j} \s_{j+1} 
    + 2 \alpha_+ \alpha_- \cos \mu \, \s_{j}^2 \s_{j+1}^2 
    \\
    & 
    + \kappa_+ \paren{ \s_{j+1} \s_{j}^2 + \s_{j+1}^2 \s_{j} }
    + \kappa_- \paren{ \s_{j} \s_{j+1}^2 + \s_{j}^2 \s_{j+1} }
    ,
    \end{aligned}
\end{align}
with the coefficients
\begin{align}
    \alpha_\pm = \frac{ \sin v_0 }{ \sin (v_0 \pm \mu) }
    ,\quad
    \kappa_\pm = \pm \frac{ \sin \mu }{ \sin (v_0 \pm \mu) } - 1
    .
\end{align}
The deformation vanishes at $v_0 = 0$, i.e. $\tilde{H}^{\mathrm{ECA}} (0) = H^{\mathrm{ECA}}$. This is a consequence of the factorisation~\eqref{eq:ECA-factorisation}, which implies that $\tilde{T} (u,0)$ is proportional to $T(u;\mu)$. In particular, the Hamiltonian~\eqref{eq:Htilde-general} with all the spectral parameters set to zero is nothing but $H^{\mathrm{ECA}}$.

For $v_0 \neq 0$, we obtain a genuinely different model, which does not commute with $H^{\mathrm{ECA}}$ and whose spectrum depends on $v_0$. It can be seen as an integrable deformation of $H^{\mathrm{ECA}}$ with a two-site unit cell, in which the parameter $v_0$ of the R matrix of non-difference form plays the role of a coupling constant~\footnote{One might also consider the sum of $\partial_u \check{R}_{j,j+1} (u,v_0;\mu) |_{u=v_0}$ over \emph{all} the sites $j$, which is invariant under the translation by one site. At $v_0 = 0$ it equals $2 H^{\mathrm{ECA}}$. For $v_0 \neq 0$, however, it is not related to the transfer matrix~\eqref{eq:T-R}, and we do not expect it to be integrable.}.

The Hamiltonian $\tilde{H}^{\mathrm{ECA}} (v_0)$ is Hermitian (in a representation where $\so_j$ and $\po_j$ are Hermitian) when $\mu$ is real and $v_0$ is imaginary, or when $\mu$ is imaginary and $v_0$ is real. In both cases $\alpha_- = \alpha_+^*$ and $\kappa_- = \kappa_+^*$. We have verified that $\tilde{H}^{\mathrm{ECA}} (v_0)$ commutes with the first higher charge generated by $\tilde{T} (u,v_0)$ on periodic chains in all the representations of Sec.~\ref{sec:rep}.

\section{Representations and $\Rep(S_3)$}
\label{sec:rep}

Any representation of $\CA$ gives rise to a representation of the Lax operator~\eqref{eq:ECA-lax} and the R matrix~\eqref{eq:ECA-R-matrix}, and thus to the medium-range Hamiltonians~\eqref{eq:ECA-Hamiltonian} and \eqref{eq:Htilde-CA}. In this section we use the relation between $\CA$ and the fusion category $\Rep(S_3)$ to obtain four of them, following the construction of Eck and Fendley~\cite{Fendley-chromatix-xxz}.

\textit{Chromatic algebra of 3-colouring and $\Rep(S_3)$.---}
The fusion category $\Rep(S_3)$ has three simple objects: the trivial representation $\mathbf{1}$, the sign representation $\epsilon$, and the two-dimensional representation $\rho$ of the symmetric group $S_3$, with fusion rules
\beq 
	\epsilon \otimes \epsilon = \mathbf{1} , \quad \rho \otimes \epsilon = \epsilon \otimes \rho = \rho, \quad \rho \otimes \rho = \mathbf{1} \oplus \epsilon \oplus \rho .
	\label{eq:fusion}
\eeq 

From the fusion category $\Rep(S_3)$, one can construct quantum lattice models by the standard construction of anyonic chains~\cite{golden-chain, Aasen2016, Aasen2020, Lootens2023, Lootens:2022avn}. The input data are the fusion category $\Rep(S_3)$ and the module category $\mathcal{M}$ over $\Rep(S_3)$~\cite{Ostrik2003, EGNO2015}. The basis states are labelled by the configurations of the simple objects of $\mathcal{M}$ on the sites of the chain, such that the objects on neighbouring sites are related by the fusion with $\rho$, and the projectors act on one site depending on its two neighbours. Most importantly, the operators acting on the Hilbert space form the \emph{bond algebra}~\cite{Cobanera:2011wn, Lootens2023, Lootens:2022avn} in terms of projectors labelled by the objects in the fusion channel. The bond algebra is directly related to $\CA$~\cite{EckFendley2023ladder},
\beq 
	\mathcal{P}^{(\mathbf{1})}_j = \frac{1}{2} \paren{ \so_j^2 + \so_j } , \quad \mathcal{P}^{(\epsilon)}_j = \frac{1}{2} \paren{ \so_j^2 - \so_j } , \quad \mathcal{P}^{(\rho)}_j = \po_j .
	\label{eq:projectors}
\eeq 

Moreover, the indecomposable module categories of $\Rep(S_3)$ are in one-to-one correspondence with the subgroups $G \subseteq S_3$ up to conjugation (the second cohomology groups of which are trivial), $\mathcal{M}_G = \Rep(G)$, on which $\Rep(S_3)$ acts by restriction. There are four of them, with $G = \mathbb{Z}_1$, $\mathbb{Z}_2$, $\mathbb{Z}_3$ and $S_3$, and they correspond to the four models found in Ref.~\cite{Fendley-chromatix-xxz} by means of orbifolds and topological defects, see Table~\ref{tab:models}.

\begin{table}
\caption{The four representations of the chromatic algebra of 3-colouring and the resulting medium-range models. The dimension of the Hilbert space is given for a periodic chain of $L$ sites. The range is the number of neighbouring sites on which the Hamiltonian density of $H^{\mathrm{ECA}}$ acts.}
\label{tab:models}
\begin{ruledtabular}
\begin{tabular}{lllcl}
$\mathcal{M}$ & Model of Ref.~\cite{Fendley-chromatix-xxz} & Hilbert space (dimension) & Range & Symmetry \\
\hline
$\Rep(\mathbb{Z}_1) = \Vect$ & XXZ chain & spins $1/2$ ($2^L$) & 4 & $S_3 \subset U(1) \rtimes \mathbb{Z}_2$ \\
$\Rep(\mathbb{Z}_2)$ & Ising zigzag ladder & spins $1/2$ ($2^L$) & 3 & $\Rep(S_3)$ \\
$\Rep(\mathbb{Z}_3)$ & three-state antiferromagnet & $s_j \neq s_{j+1}$ ($2^L + 2 (-1)^L$) & 5 & $S_3$ \\
$\Rep(S_3)$ & Rydberg-blockade ladder & blockade ($2^L + (-1)^L$) & 5 & $\Rep(S_3)$ \\
\end{tabular}
\end{ruledtabular}
\end{table}

The symmetry of the model associated with $\mathcal{M}$ is described by the dual category of $\Rep(S_3)$ with respect to $\mathcal{M}$~\cite{EGNO2015, Lootens2023, Lootens:2022avn}. It is the group $S_3$ for $G = \mathbb{Z}_1$ and $\mathbb{Z}_3$, while for $G = \mathbb{Z}_2$ and $S_3$ it is $\Rep(S_3)$ itself, whose object $\rho$ gives rise to a non-invertible symmetry operator~\cite{BhardwajTachikawa2018, Chang2019, SchaferNameki2024}.
The symmetry operators, as well as the non-invertible duality maps between the four models, were constructed explicitly in Ref.~\cite{Fendley-chromatix-xxz}. Importantly, they commute with (respectively intertwine) the individual generators $\so_j$ and $\po_j$, and hence any Hamiltonian composed of them. Therefore, the medium-range models below inherit the symmetries listed in Table~\ref{tab:models}, and they are mapped to each other by the non-invertible duality maps of Ref.~\cite{Fendley-chromatix-xxz}.

We now present the Hamiltonian $H^{\mathrm{ECA}}$ in the four representations. The Hamiltonian $\tilde{H}^{\mathrm{ECA}} (v_0)$ is obtained in the same way from Eqs.~\eqref{eq:Htilde-CA} and \eqref{eq:delta-h}.

\textit{XXZ representation.---}
One of the physical representations can be derived by representing the generators of ECA by the XXZ representation of $\CA$~\eqref{eq:CA-rep-xxz}, in which
\begin{align}
    \s_j = \frac{1}{4} \paren{ 1 + Z_{j-1} Z_{j+2} } \paren{ X_{j} X_{j+1} + Y_{j} Y_{j+1} }
    ,\quad
    \s_j^2 = \frac{1}{4} \paren{ 1 + Z_{j-1} Z_{j+2} } \paren{ 1 - Z_{j} Z_{j+1} }
    .
\end{align}
The Hamiltonian~\eqref{eq:ECA-Hamiltonian} becomes
\begin{align}
    \label{eq:H-xxz}
    H^{\mathrm{ECA}}_{\mathrm{XXZ}} 
    = 
    \frac{1}{4} \sum_{j=1}^{L} \paren{ 1 + Z_{j-1} Z_{j+2} } \bck{ X_{j} X_{j+1} + Y_{j} Y_{j+1} + \Delta \paren{ 1 - Z_{j} Z_{j+1} } }
    ,
\end{align}
with a four-site interaction. 
The special case of this representation $\Delta = 0$ corresponds to the folded XXZ chain~\cite{Lenart-folded1-2021, Lenart-folded2-2021, Pozsgay-folded-xxz-2021}, which is equal to the special point of the Bariev chain~\cite{Bariev1991-original,Bariev1991-XY,Bariev1993} when considering the Kramers-Wannier duality~\cite{KramersWannier1941}. The conserved charge~\eqref{eq:NP} is $N_P = \frac{1}{2} \sum_j ( 1 + Z_j Z_{j+1} )$, i.e. the number of domain walls is conserved, which is a hallmark of the folded XXZ chain.

\textit{Ising zigzag representation.---}
In the representation associated with $\Rep(\mathbb{Z}_2)$, the generators act on a chain of spins $1/2$ as~\cite{Fendley-chromatix-xxz}
\begin{align}
    \so_j^{\mathrm{Izz}} = \frac{1}{2} \sigma^z_{j-1} \paren{ 1 + \sigma^x_j } \sigma^z_{j+1}
    ,\quad
    \po_j^{\mathrm{Izz}} = \frac{1}{2} \paren{ 1 - \sigma^x_j }
    ,
\end{align}
where $\sigma^{\alpha}_j$ are the Pauli matrices acting on $j$-th site. The Hamiltonian $H^{\mathrm{CA}}$ describes two Ising chains coupled in a zigzag fashion. The Kramers-Wannier duality maps the XXZ representation to the zigzag one~\cite{Fendley-chromatix-xxz}. 
Since $\po_j$ acts on a single site, the generators of ECA act on three sites only,
\begin{align}
    \s_j = \frac{1}{4} \paren{ 1 + \sigma^x_{j} } \paren{ \sigma^z_{j-1} \sigma^z_{j+1} + \sigma^y_{j-1} \sigma^y_{j+1} }
    ,\quad
    \s_j^2 = \frac{1}{4} \paren{ 1 + \sigma^x_{j} } \paren{ 1 - \sigma^x_{j-1} \sigma^x_{j+1} }
    .
\end{align}
After the rotation $( \sigma^x , \sigma^y , \sigma^z ) \to ( \sigma^z , - \sigma^y , \sigma^x )$, the Hamiltonian~\eqref{eq:ECA-Hamiltonian} becomes
\begin{align}
    \label{eq:H-zigzag}
    H^{\mathrm{ECA}}_{\mathrm{zig}} 
    = 
    \sum_{j=1}^{L} n_j \bck{ \sigma^+_{j-1} \sigma^-_{j+1} + \sigma^-_{j-1} \sigma^+_{j+1} + \frac{\Delta}{2} \paren{ 1 - \sigma^z_{j-1} \sigma^z_{j+1} } }
    ,\quad
    n_j = \frac{1}{2} \paren{ 1 + \sigma^z_j }
    ,
\end{align}
i.e. the XXZ interaction between the next-nearest neighbours that is switched on by the spin in between. 
This is the hard-rod deformed XXZ model of Refs.~\cite{Pozsgay-medium-2021, Pozsgay-hardrod-2021}, cf. also Ref.~\cite{AlcarazBariev1999}, which reduces to the Bariev model~\cite{Bariev1991-original,Bariev1991-XY,Bariev1993} with the coupling $|U| = 1$ at $\Delta = 0$. In this representation, the Lax operator~\eqref{eq:ECA-lax} and the R matrix~\eqref{eq:ECA-R-matrix} act on three and four sites, respectively, and they fit into the framework of Ref.~\cite{Pozsgay-medium-2021}.

The conserved charge~\eqref{eq:NP} is the magnetisation, $N_P = \sum_j ( 1 - n_j )$. Besides, our construction shows that the hard-rod deformed XXZ model possesses a non-invertible $\Rep(S_3)$ symmetry, inherited from the Ising zigzag ladder~\cite{Fendley-chromatix-xxz}.

\textit{Three-state antiferromagnet.---} 
In the representation associated with $\Rep(\mathbb{Z}_3)$, each site hosts a three-state spin $s_j \in \{ A, B, C \}$, subject to the constraint $s_j \neq s_{j+1}$. The generators are~\cite{Fendley-chromatix-xxz}
\begin{align}
    \so_j^{(3)} = \delta_{s_{j-1}, s_{j+1}} F_j
    ,\quad
    \po_j^{(3)} = 1 - \delta_{s_{j-1}, s_{j+1}}
    ,
\end{align}
where $F_j$ flips the spin $s_j$ to the unique state that differs from both $s_j$ and $s_{j-1} = s_{j+1}$. 
The Hamiltonian~\eqref{eq:ECA-Hamiltonian} becomes
\begin{align}
    \label{eq:H-af3}
    H^{\mathrm{ECA}}_{3} 
    = 
    \sum_{j=1}^{L} \paren{ 1 - \delta_{s_{j-2}, s_{j+2}} } \delta_{s_{j-1}, s_{j+1}} \paren{ F_j + \Delta }
    ,
\end{align}
with a five-site interaction: the spin $s_j$ is flipped only if its nearest neighbours coincide and its next-nearest neighbours differ. The model is invariant under the permutations of the three states, which form the group $S_3$.

\textit{Rydberg-blockade ladder.---} 
In the representation associated with the regular module category $\Rep(S_3)$, each rung $j$ of a two-leg ladder is either empty, or occupied by a hard-core boson on the top or on the bottom leg, annihilated by $t_j$ and $b_j$, respectively. Two neighbouring rungs cannot be occupied simultaneously, due to Rydberg blockade mechanism. Following Ref.~\cite{Fendley-chromatix-xxz}, we denote
\begin{align}
    d_{j,\pm} = \frac{ t_j \pm b_j }{ \sqrt{2} }
    ,\quad
    n^{\pm}_j = d_{j,\pm}^\dagger d_{j,\pm}^{\phantom{\dagger}}
    ,\quad
    n^{(e)}_j = 1 - n^{+}_j - n^{-}_j
    ,\quad
    \varsigma_j = d_{j,+}^\dagger d_{j,-}^{\phantom{\dagger}} + d_{j,-}^\dagger d_{j,+}^{\phantom{\dagger}}
    ,
\end{align}
where the operators $d_{j,\pm}$ are understood to be projected onto the states in which the rungs $j \pm 1$ are empty. The generators are
\begin{align}
    \so_j^{\mathrm{IRL}} = d_{j,+}^{\phantom{\dagger}} + d_{j,+}^\dagger + \varsigma_{j-1} \varsigma_{j+1}
    ,\quad
    \po_j^{\mathrm{IRL}} = n^{-}_j + \paren{ n^{(e)}_{j-1} - n^{(e)}_{j+1} }^2
    ,
\end{align}
and $H^{\mathrm{CA}}$ is the Hamiltonian of the integrable Rydberg-blockade ladder~\cite{Fendley-chromatix-xxz, EckFendley2023ladder}. 
The Hamiltonian~\eqref{eq:ECA-Hamiltonian} becomes
\begin{align}
    \label{eq:H-irl}
    \begin{aligned}
    H^{\mathrm{ECA}}_{\mathrm{IRL}} 
    = 
    \sum_{j=1}^{L} \Big[ &
    \paren{ n^{(e)}_{j-2} - n^{(e)}_{j+2} }^2 \paren{ d_{j,+}^{\phantom{\dagger}} + d_{j,+}^\dagger + \Delta \, n^{(e)}_{j-1} \paren{ 1 - n^{-}_j } n^{(e)}_{j+1} }
    \\
    & + 
    \paren{ n^{-}_{j-1} - n^{-}_{j+1} }^2 \paren{ \varsigma_{j-1} \varsigma_{j+1} + \Delta \paren{ 1 - n^{(e)}_{j-1} } \paren{ 1 - n^{(e)}_{j+1} } }
    \Big]
    ,
    \end{aligned}
\end{align}
with a five-rung interaction. 
The first line describes the creation and the annihilation of a boson in the symmetric state on the rung $j$, which is allowed only if exactly one of the rungs $j \pm 2$ is occupied. The second line exchanges the bosons in the symmetric and the antisymmetric states on the rungs $j-1$ and $j+1$. 
Similar to the hard-rod deformed XXZ model, the model~\eqref{eq:H-irl} possesses a non-invertible $\Rep(S_3)$ symmetry.

\emph{Remark.} As noted in~\cite{YM_global_symm}, the Hamiltonians \eqref{eq:H-zigzag} and \eqref{eq:H-irl} both possess a continuous non-invertible symmetry, whose fusion rules are those of the double-coset ring $\mathbb{Z}\left[\mathbb{Z}_2 \backslash O(2) / \mathbb{Z}_2\right]$. The discrete non-invertible symmetry $\Rep(S_3)$ is contained in this continuous symmetry.

To the best of our knowledge, the models~\eqref{eq:H-af3} and \eqref{eq:H-irl}, as well as the deformations $\tilde{H}^{\mathrm{ECA}} (v_0)$ of the four models, are new. Let us finally remark on a property shared by the four models. Since $\s_j$ acts non-trivially only when $\po_j = 0$ and $\po_{j-1} \neq \po_{j+1}$, and exchanges the values of $\po_{j-1}$ and $\po_{j+1}$, the configuration of the eigenvalues of the commuting projectors $\po_j$ evolves, in any representation, as that of hard-core particles hopping by two sites across an empty site. This is the kinetic constraint responsible for the Hilbert space fragmentation of the folded XXZ chain~\cite{Lenart-folded1-2021, Pozsgay-folded-xxz-2021}, which we therefore expect to be present in all four models.

\section{Summary and outlook}
\label{sec:summary}
We have proposed an extension of Jones' Baxterisation to medium-range solvable models. It is based on the extended Yang-Baxter equation~\eqref{eq:EYBE} and the RLL relation~\eqref{eq:RLL}, formulated for the elements of an algebra that commute at distance three or larger. As in Jones' construction, they lead to transfer matrices in involution up to boundary terms, and to the medium-range Hamiltonians~\eqref{eq:H-general} and \eqref{eq:Htilde-general}, which we conjecture to be integrable in any representation of the algebra.

We have found a solution in the chromatic algebra of 3-colouring, given by the Lax operator~\eqref{eq:ECA-lax} and the R matrix of non-difference form~\eqref{eq:ECA-R-matrix}. The relation between the chromatic algebra of 3-colouring and $\Rep(S_3)$ yields four representations. They unify the folded XXZ model and the hard-rod deformed XXZ model with two new models, defined on a three-state antiferromagnet and on a Rydberg-blockade ladder. The four models are related by non-invertible duality maps, and the hard-rod deformed XXZ model and the Rydberg-blockade ladder possess a non-invertible $\Rep(S_3)$ symmetry.

There are several open questions. 
First, the spectra of the new models remain to be found. The Bethe ansatz solution of the hard-rod deformed XXZ model is known~\cite{Pozsgay-hardrod-2021}, and the relation between the four representations is reminiscent of the vertex-IRF correspondence~\cite{Baxter1973-i, Baxter1973-ii, Baxter1973-iii, Pasquier1988}. It is desirable to formulate the diagonalisation of the transfer matrices~\eqref{eq:T-lax} and \eqref{eq:T-R} at the level of the algebra, which would give access to the thermodynamics and the generalised hydrodynamics of all four models at once. 
Second, the R matrix~\eqref{eq:ECA-R-matrix} is not of difference form. Since the Hamiltonian~\eqref{eq:H-zigzag} at $\Delta = 0$ is a special point of the Bariev model, it would be interesting to clarify the relation to the R matrix of the Bariev model~\cite{Zhou1996, Shiroishi1997, Bariev1993}, and whether the latter with generic coupling admits an algebraic formulation of the same kind. 
Third, we have only presented one solution to the extended YBE. It is natural to search for solutions in other Hecke-type algebras, and to classify the solutions in a given algebra. The construction can also be generalised to elements of the algebra that commute only at larger distances, corresponding to interactions of longer range. 
Finally, the unitarity~\eqref{eq:ECA-unitarity} suggests using the R matrix as a quantum gate, which would lead to integrable quantum circuits and cellular automata with non-invertible symmetries~\cite{Vanicat2018, Miao2024floquet}. The extension to open boundary conditions~\cite{Sklyanin1988} is another interesting direction.

\section*{Acknowledgement}
Y.M. has received support under the Major Research Program of PSL Research University ``Statistical Physics and Mathematics'' launched by PSL Research University and implemented by ANR with the references ANR-10-IDEX-0001. The authors obtained the results of the paper without the assistance of AI. We acknowledge the use of Claude Opus 5.5 and Fable 5.2 to assist in proofreading the manuscript.

\appendix

\section{Relations of the extended chromatic algebra}
\label{app:relations}

In this appendix we derive the relations~\eqref{eq:ECA-def-1}--\eqref{eq:ECA-def-4} from the relations of $\CA$. It is convenient to introduce $K_j = 2 \po_j - 1$. From Eqs.~\eqref{eq:CA-def1}--\eqref{eq:CA-PS} we have
\begin{align}
    \label{eq:K-rel}
    K_j^2 = 1
    ,\quad
    \bck{ K_j , K_l } = 0
    ,\quad
    K_j \so_j = \so_j K_j = - \so_j
    ,\quad
    K_{j \pm 1} \so_j = - \so_j K_{j \pm 1}
    ,
\end{align}
while $K_l$ commutes with $\so_j$ for $|j - l| \geq 2$. Thus, the generators~\eqref{eq:ECA-embedding} are expressed as
\begin{align}
    \label{eq:ECA-K}
    \s_j = \Pi_j \so_j = \so_j \Pi_j
    ,\quad
    \w_j = \frac{1}{2} \paren{ 1 + K_{j-1} K_j K_{j+1} }
    ,\quad
    \p_j = \frac{1}{2} \paren{ 1 + K_j } \paren{ 1 - \Pi_j }
    ,
\end{align}
with the projector $\Pi_j = ( \po_{j-1} - \po_{j+1} )^2 = \frac{1}{2} ( 1 - K_{j-1} K_{j+1} )$. The first equality follows from $[ \po_{j-1} , \so_j ] = K_{j-1} \so_j$ and $[ K_{j-1} \so_j , \po_{j+1} ] = - K_{j-1} K_{j+1} \so_j$, and $\Pi_j$ commutes with $\so_j$ since $\so_j$ anticommutes with both $K_{j-1}$ and $K_{j+1}$.

The relations~\eqref{eq:ECA-def-1} follow immediately from Eq.~\eqref{eq:ECA-K}, e.g.
$\s_j^2 = \Pi_j ( 1 - \po_j ) = \frac{1}{2} \Pi_j ( 1 - K_j ) = \w_j - \p_j$, where we used $K_j \so_j = - \so_j$, which also implies $\w_j \so_j = \Pi_j \so_j$.

The elements $\p_j$ and $\w_j$ are polynomials in $K_l$ and hence commute with each other. For $|j-l| > 2$, the elements $\s_j$ and $A_l \in \{ \s_l , \p_l , \w_l \}$ are built from generators of $\CA$ whose indices differ by more than one, except for pairs of $K$'s, which commute. This proves Eq.~\eqref{eq:ECA-def-3}.

For $d \in \{ 1, 2 \}$, the product $K_{j-1} K_j K_{j+1}$ contains exactly one factor that anticommutes with $\s_{j \pm d}$, namely $K_{j}$ for $d=1$ and $K_{j \pm 1}$ for $d = 2$, while the remaining factors commute with it. Hence $K_{j-1} K_j K_{j+1}$ anticommutes with $\s_{j \pm d}$, which is Eq.~\eqref{eq:ECA-def-4}.

Finally, we prove Eq.~\eqref{eq:ECA-def-2}. The case $A_j = \w_j$ follows from Eq.~\eqref{eq:ECA-def-4}, and the case $A_j = \p_j$ is proved in the same way, using $\p_j \s_{j \pm 1} = \s_{j \pm 1} \frac{1}{2} ( 1 - K_j ) ( 1 - \Pi_j )$ and $\p_j \s_{j \pm 2} = \s_{j \pm 2} \frac{1}{2} ( 1 + K_j ) \Pi_j$, which are annihilated by $\p_j$ from the right. For $A_j = \s_j$ and $d = 1$, we move all the projectors to the left using Eq.~\eqref{eq:K-rel}, and the result is proportional to $\so_j \so_{j \pm 1} \so_j = 0$. For $A_j = \s_j$ and $d = 2$, the generators $\so_j$ and $\so_{j \pm 2}$ commute, but $\so_{j \pm 2}$ anticommutes with $K_{j \pm 1}$, so that $\Pi_j \s_{j \pm 2} = \s_{j \pm 2} ( 1 - \Pi_j )$ and
\begin{align}
    \s_j \s_{j \pm 2} \s_j = \so_j \Pi_j \s_{j \pm 2} \Pi_j \so_j = \so_j \s_{j \pm 2} \paren{ 1 - \Pi_j } \Pi_j \so_j = 0 .
\end{align}

The same reasoning shows that $N_P = \frac{1}{2} \sum_j ( 1 + K_j )$ commutes with $\s_j$: the generator $\so_j$ flips the signs of $K_{j-1}$ and $K_{j+1}$ and preserves the other $K_l$, and $\Pi_j$ projects onto $K_{j-1} + K_{j+1} = 0$.

\bibliography{refs_arXiv_sub}

\end{document}